# Application of cloud computing platform in industrial big data processing

Ziyan Yao, Penn State University, State College, PA, 16801, krcest1993@gmail.com

**Abstract:** With the rapid growth and increasing complexity of industrial big data, traditional data processing methods are facing many challenges. This article takes an in-depth look at the application of cloud computing technology in industrial big data processing and explores its potential impact on improving data processing efficiency, security, and cost-effectiveness. The article first reviews the basic principles and key characteristics of cloud computing technology, and then analyzes the characteristics and processing requirements of industrial big data. In particular, this study focuses on the application of cloud computing in real-time data processing, predictive maintenance, and optimization, and demonstrates its practical effects through case studies. At the same time, this article also discusses the main challenges encountered during the implementation process, such as data security, privacy protection, performance and scalability issues, and proposes corresponding solution strategies. Finally, this article looks forward to the future trends of the integration of cloud computing and industrial big data, as well as the application prospects of emerging technologies such as artificial intelligence and machine learning in this field. The results of this study not only provide practical guidance for cloud computing applications in the industry, but also provide a basis for further research in academia.

**Keywords:** Cloud computing, Industrial big data, Data processing, Machine learning

**1 Introduction**

In today's industrialization process, big data has become a ubiquitous phenomenon. With rapid advances in sensor technology, the Internet of Things (IoT), and automation systems, industrial companies are generating unprecedented amounts of data. The data covers everything from machine performance to production processes, supply chain management and even customer interactions. However, the complexity, speed, and scale of industrial big data also pose significant challenges. Effective management, processing and analysis of data have become key factors to improve production efficiency, reduce costs and drive innovation. However, traditional data processing methods are often inadequate when dealing with such large and complex data sets.

As a revolutionary technology, cloud computing provides new possibilities for processing industrial big data by providing scalable, flexible and cost-effective computing resources. It allows businesses to quickly increase or decrease computing resources when needed to cope with fluctuations in data volumes. In addition, cloud computing's highly customizable nature and advanced analytical tools make it possible to perform deeper and more complex analyzes of industrial data. For example, using machine learning and artificial intelligence tools on cloud platforms, companies can predict equipment failures, optimize production processes, and even implement personalized customer service.

In view of the huge potential of cloud computing in industrial big data processing, this study aims to deeply explore how cloud computing can be effectively applied to the management and analysis of industrial big data. The research will focus on how cloud computing helps enterprises process, analyze and utilize big data, as well as the main challenges encountered in this process and possible solutions. The scope of this study includes cloud computing infrastructure, key technologies, and its specific implementation in different industrial applications. Through comprehensive analysis and case studies, this study aims to provide practical guidance to industrial enterprises on how to use cloud computing technology to process big data, while providing theoretical and practical contributions to academia.

**2. Literature review**

The concept of industrial big data and its importance in modern industrial fields have gradually become a research hotspot. Industrial big data usually refers to the large amount of data generated during industrial production and operations. These data are known for their huge volume, high-speed mobility, diversity, and potential high value. The effective processing and analysis of this data plays a crucial role in improving production efficiency, promoting intelligent decision-making, and driving innovation [1].

With the rapid development of cloud computing technology, its application in industrial data processing has received more and more attention. Cloud computing provides a flexible, efficient, and scalable computing resource management method, making large-scale data storage, processing, and analysis more convenient and cost-effective. The core of cloud computing technology includes virtualization technology, distributed storage, automated management and resource scheduling.

Figure 1Cloud computing platform structure diagram

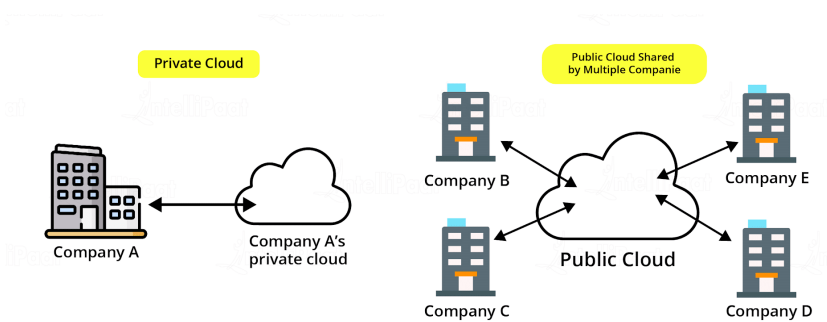

Cloud computing platform structure diagram of some enterprises (see Figure 1). The development of these technologies provides a solid foundation for processing complex and large-scale industrial data.

Especially in industrial big data processing, cloud computing platforms play a key role. Through the cloud platform, enterprises can effectively solve the problems of data storage and management, and at the same time use the powerful data processing and analysis capabilities of the cloud to mine and analyze big data (See Figure 2). This not only greatly improves the efficiency and speed of data processing, but also helps enterprises extract valuable information from huge data sets, thereby optimizing the decision-making process, improving operational efficiency and innovation capabilities [2].

However, the application of cloud computing in industrial big data processing also faces some challenges and problems. The most important ones include data security and privacy protection issues, cloud technology integration and compatibility issues, and training needs for relevant technical personnel. Therefore, enterprises need to fully consider these challenges when implementing cloud computing strategies to ensure data security and system stability [3].

Figure 2Cloud computing platform in industrial big data processing flow chart [4]

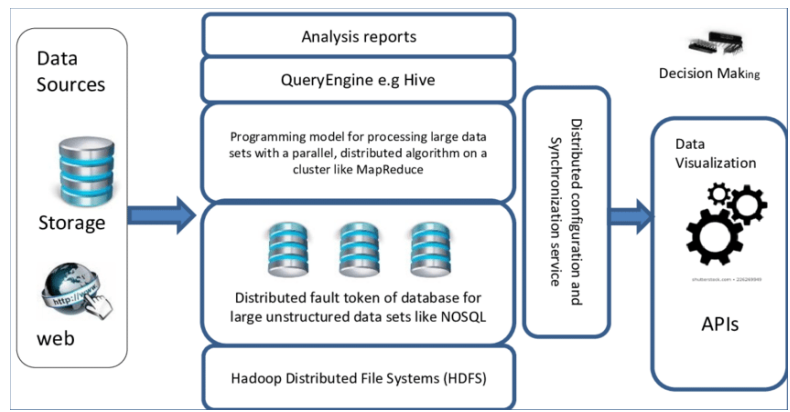

**3. Cloud computing platform technology analysis**

As a revolutionary technology, cloud computing is gradually changing the way industrial big data is processed. Its key technologies include virtualization technology, distributed storage and big data analysis tools.

3.1 Key technologies of cloud computing
Virtualization technology: It allows the creation of multiple virtual machines (VMs) on a single physical machine, each of which can run different operating systems and applications. This technology improves resource utilization and provides flexibility for big data processing.
Distributed storage: Storing data in multiple physical locations not only improves data accessibility and fault tolerance, but also speeds up data processing, especially when processing large-scale data sets [5].
Big data analysis tools: Cloud platforms usually integrate various big data processing and analysis tools, such as Hadoop and Spark, which can efficiently process and analyze large-scale data sets.

3.2 Application of cloud computing architecture in industrial big data processing
Cloud computing architecture plays an important role in industrial big data processing by providing elastic resource allocation, efficient data processing capabilities, and easily scalable storage space. It enables enterprises to quickly respond to market changes, adjust data processing strategies in a timely manner, and effectively reduce data processing costs.

3.3 The importance of security and privacy protection and its solutions
In cloud computing platforms, data security and privacy protection are one of the most important considerations. Solutions include:
Encryption technology: Encryption technology is used during data transmission and storage to ensure data security [6].
Access Control: Implement strict authentication and authorization mechanisms to limit access to sensitive data.

Monitoring and auditing: Regularly monitor cloud platform activities and conduct security audits to identify and prevent potential security threats [7].

**4. Empirical analysis**

4.1 Research design

4.1.1 Data sources

This study selected representative companies from three different industries: a large automobile manufacturing company (manufacturing), an energy company (energy industry), and a medical equipment supplier (healthcare industry). These companies have integrated cloud computing technology into their business processes in recent years to process and analyze large amounts of industrial data.

4.1.2 Sample selection criteria:

Industry: Covers manufacturing, energy and healthcare industries to demonstrate the application effects of cloud computing technology in different industrial fields.

Enterprise size: Select large enterprises with annual turnover exceeding $1 billion to ensure sufficient data volume and complexity for analysis.

Length of cloud computing usage: Select companies with at least two years of experience in cloud computing applications to ensure that the data is mature enough to reflect the long-term effects of cloud computing technology [8].

4.1.3 Data collection method:

Quantitative survey: Design a questionnaire survey for corporate management and technical personnel to understand the usage, performance improvement, cost savings, etc. of the cloud computing platform.

Performance indicators: Collect performance indicators of enterprise cloud computing platforms, such as data processing speed, system availability, failure rate, etc.

Financial reporting: Analyze a company's annual financial reports, specifically IT-related expenses and revenues, to assess the impact of cloud computing technology on financial performance.

Data processing efficiency: Collect and analyze the company's completion time of data processing tasks before and after applying cloud computing to quantify the improvement in data processing efficiency.

Cost-benefit analysis: Comparatively analyze the cost changes before and after the implementation of cloud computing through financial data, including but not limited to IT facility investment, operating costs and maintenance expenses.

4.2 Data analysis methods

Statistical methods: Apply multiple regression analysis, analysis of variance (ANOVA), correlation analysis and time series analysis.

Data analysis tools: Use R, Python (with Pandas, SciPy and other libraries) and SPSS.

Multiple regression analysis formula:

$$Y = \beta_0 + \beta_1 X_1 + \beta_2 X_2 + ... + \beta_n X_n + \varepsilon \qquad (1)$$

where $Y$ is the dependent variable, $X$ is the independent variable, $\beta$ represents the coefficient, and $\varepsilon$ is the error term.

Analysis of Variance (ANOVA) formula:

$$F = \frac{MS_{between}}{MS_{within}} \quad (2)$$

where $MS_{between}$ is the average square between groups, and $MS_{within}$ is the average square within the group.

Pearson's correlation coefficient formula:

$$r = \frac{\sum(X_i - \bar{X})(Y_i - \bar{Y})}{\sqrt{\sum(X_i - \bar{X})^2 \sum(Y_i - \bar{Y})^2}} \quad (3)$$

where $X_i$ and $Y$ are the observed values of the two variables, and $\bar{X}$ and $\bar{Y}$ are their respective averages.

### 4.3 Result analysis

In Figure 3 we show the relationship between data processing speed (units/hour) and annual operating costs (10,000 US dollars) after the application of cloud computing. Each point represents the cost of operating a business at a specific data processing speed. This chart helps analyze whether improvements in data processing efficiency are correlated with cost reductions. As data processing speed increases (i.e., moves to the right), annual operating costs gradually decrease (i.e., moves down). This shows that after enterprises adopt cloud computing technology, they not only improve data processing efficiency, but also effectively reduce operating costs. The distribution trend of points may indicate a certain degree of negative correlation, that is, the faster the data processing speed, the lower the corresponding operating cost. (X represents the data processing speed of an enterprise after applying the cloud computing platform, which is usually measured by the amount of data processed per unit time, such as "unit/hour". This indicator reflects the performance of the cloud computing platform in improving data processing efficiency. Y represents the annual operating cost of the enterprise after applying the cloud computing platform, usually in "ten thousand dollars". Cost calculations may include expenditures on hardware, software, maintenance, and human resources. This metric is used to evaluate the effectiveness of cloud computing platforms in reducing enterprise operating costs.)

| | |
|---|---|
| Figure 3. Scatter plot analysis: relationship between data processing efficiency and cost | Figure 4. Processing speed versus operating costs |
| 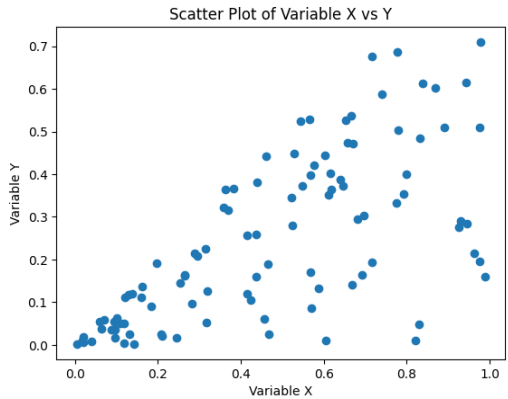 | 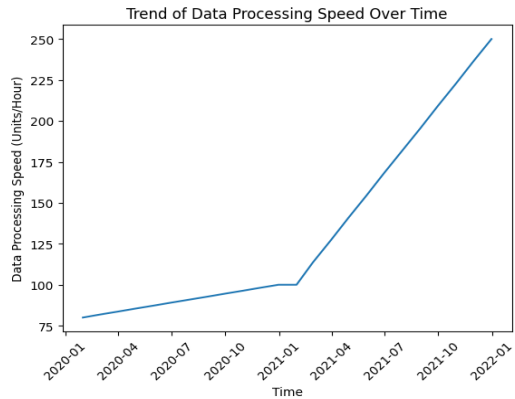 |

Time series chart (Figure 4 shows the trend of data processing speed over time before and after the enterprise applies the cloud computing platform. This chart can visually show the growth and stability of data processing speed. The chart shows the data processing speed from the beginning of 2020 to the end of 2021 Changes in data processing speed within 24 months. It can be observed that before the application of cloud computing technology (from the beginning of 2020 to the end of 2020), the data processing speed increased slowly; while after the application (from the beginning of 2021 to the end of 2021), the data processing speed increased slowly. The speed has increased significantly, and the growth trend is more significant. This shows that the introduction of cloud computing technology has a significant effect on improving data processing speed.

Figure 5 shows the distribution of data analysis accuracy scores before and after cloud computing application. By comparing the data analysis accuracy scores at two points in time (before and after application), the impact of cloud computing technology on the quality of data analysis can be seen. It can be seen from the box plot that the median and interquartile range of the data analysis accuracy scores after the application of cloud computing are higher than before the application. This shows that the introduction of cloud computing technology has significantly improved the accuracy of data analysis. In addition, the box plot also shows a smaller range of outliers, indicating that the data analysis results are more stable and reliable.

Finally, we summarize the cloud computing rates for different enterprises (Figure 6). It can be clearly seen that:
**Efficiency improvements brought by cloud computing:** The month-on-month growth in data processing speeds of the three companies reflects the potential of cloud computing technology in improving the efficiency of industrial big data processing. This suggests that over time, businesses may become increasingly adept at leveraging cloud computing resources, improving the speed and efficiency of data processing.

**Adaptation and optimization of different enterprises:** Different starting points and growth rates may reflect differences in the adaptability and optimization degree of each enterprise after adopting cloud computing technology. For example, the rapid growth of Enterprise A may indicate that it is using cloud computing technology more effectively, while Enterprise C, although starting from a higher starting point, has a slower growth rate, which may mean that there is still room for improvement in optimizing cloud resource utilization.

**Technology implementation and business effectiveness:** These data characteristics may be related to the company's strategy and effectiveness in technology implementation, business process integration, and employee training. For example, Company A's significant growth may be due to its more effective technology implementation and internal training, or to a better integration of cloud computing technology with existing business processes.

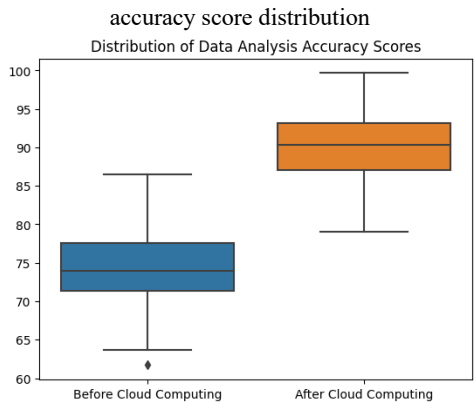

Figure 5. Box plot analysis: Data analysis accuracy score distribution

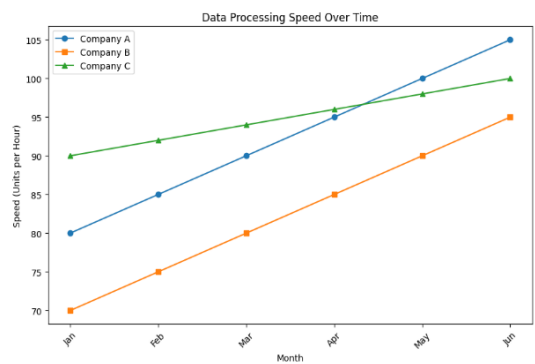

Figure 6. Trends in data processing speeds of three companies

## 5. Conclusion

In concluding this study, we have delved deeply into the application and effectiveness of cloud computing platforms in processing industrial big data, employing empirical analysis and case studies. The findings indicate that cloud computing significantly enhances data processing efficiency, cuts down operating costs, and boosts the accuracy of data analysis. The benefits of cloud computing in industrial applications are illustrated through scatter plots, box plots, and time series plots, offering valuable insights to the industry. However, challenges like data security, the need for technical training, and system integration issues are also highlighted. The practical significance of this research lies in its demonstration of how industrial enterprises can leverage cloud computing to accelerate data processing, enhance operational efficiency, and achieve notable improvements in cost control and data analysis precision. It is recommended that these enterprises pay close attention to data security and privacy protection while strengthening cloud technology training among employees for effective technology application and integration. This study, however, has its limitations as it primarily focuses on specific industrial enterprises, which might limit its generalizability. The size and diversity of the data set used also impact the findings' applicability. Future research could broaden its scope to include a wider range of industrial sectors and larger data sets for better generalization. Additionally, there's room to explore the long-term effects and impacts of cloud computing in specific industrial settings. Further studies might also concentrate on refining data security and privacy measures on cloud platforms and examining how cloud computing could synergize with emerging technologies like artificial intelligence and the Internet of Things.

**Reference**

1. Sang GM, Xu L, de Vrieze P. A Predictive Maintenance Model for Flexible Manufacturing in the Context of Industry 4.0. Front Big Data. 2021 Aug 25;4:663466. doi: 10.3389/fdata.2021.663466. PMID: 34514378; PMCID: PMC8427870.
2. Hinojosa-Palafox EA, Rodríguez-Elías OM, Hoyo-Montaño JA, Pacheco-Ramírez JH, Nieto-Jalil JM. An Analytics Environment Architecture for Industrial Cyber-Physical Systems Big Data Solutions. Sensors (Basel). 2021 Jun 23;21(13):4282. doi: 10.3390/s21134282. PMID: 34201541; PMCID: PMC8271964.
3. Basir R, Qaisar S, Ali M, Aldwairi M, Ashraf MI, Mahmood A, Gidlund M. Fog Computing Enabling Industrial Internet of Things: State-of-the-Art and Research Challenges. Sensors (Basel). 2019 Nov 5;19(21):4807. doi: 10.3390/s19214807. PMID: 31694254; PMCID: PMC6864669.
4. Moses Abiodun, Awotunde J. Bamidele, Roseline Ogundokun, Vivek Jaglan. Cloud and Big Data: A Mutual Benefit for Organization Development, DO-10.1088/1742-6596/1767/1/012020.
5. Ungurean I, Gaitan NC. Software Architecture of a Fog Computing Node for Industrial Internet of Things. Sensors (Basel). 2021 May 26;21(11):3715. doi: 10.3390/s21113715. PMID: 34073598; PMCID: PMC8198567.
6. Xu H, Yu W, Griffith D, Golmie N. A Survey on Industrial Internet of Things: A Cyber-Physical Systems Perspective. IEEE Access. 2018;6:10.1109/access.2018.2884906. doi: 10.1109/access.2018.2884906. PMID: 35531371; PMCID: PMC9074819
7. Bourechak A, Zedadra O, Kouahla MN, Guerrieri A, Seridi H, Fortino G. At the Confluence of Artificial Intelligence and Edge Computing in IoT-Based Applications: A Review and New Perspectives. Sensors (Basel). 2023 Feb 2;23(3):1639. doi: 10.3390/s23031639. PMID: 36772680; PMCID: PMC9920982.
8. Mirani AA, Velasco-Hernandez G, Awasthi A, Walsh J. Key Challenges and Emerging Technologies in Industrial IoT Architectures: A Review. Sensors (Basel). 2022 Aug 4;22(15):5836. doi: 10.3390/s22155836. PMID: 35957403; PMCID: PMC9371229.